\documentclass[a4paper,11pt]{article}

\usepackage{jheppub}
\usepackage[dvipsnames]{xcolor}
\usepackage{lineno}
\usepackage{diagbox}
\usepackage{amsfonts,amsmath}
\usepackage{bbold}
\usepackage{tikz}
\usepackage{booktabs}
\usepackage{verbatim}
\usepackage{hhline}
\usetikzlibrary{positioning}
\usepackage[capitalise]{cleveref}
\usepackage{hyperref}
\usepackage{colortbl}
\hypersetup{colorlinks, urlcolor=BlueViolet, citecolor=Plum, linkcolor=PineGreen}

\newcommand{\mX}{\mathcal{X}}
\newcommand{\mN}{\mathcal{N}}

\title{Classification of three-family flavoured DFSZ axion models that have no domain wall problem}

\author[a]{Peter Cox,}
\author[a]{Matthew J. Dolan,}
\author[a]{Maaz Hayat\footnote{Corresponding author},}
\author[a,b]{Andrea Thamm,}
\author[a]{Raymond R. Volkas}

\affiliation[a]{ARC Centre of Excellence for Dark Matter Particle Physics,\\
School of Physics, The University of Melbourne,\\
Victoria 3010 Australia}
\affiliation[b]{Department of Physics, University of Massachusetts Amherst, MA 01003, USA}

\emailAdd{peter.cox@unimelb.edu.au}
\emailAdd{dolan@unimelb.edu.au}
\emailAdd{maazh@student.unimelb.edu.au}
\emailAdd{athamm@umass.edu}
\emailAdd{raymondv@unimelb.edu.au}

\abstract{We provide an exhaustive classification of three-family DFSZ axion models that have no cosmological domain wall problem. This result is obtained by making the Peccei-Quinn symmetry flavour dependent in certain specific ways, thus reinforcing a possible connection between the strong CP problem and the flavour puzzle. Known DFSZ flavour variants such as the top-specific model emerge as special cases. Key features of the phenomenology of these models are briefly discussed.}

\begin{document}
\maketitle
\flushbottom

\section{Introduction}
\label{sec:intro}

The strong CP problem is an important motivation for physics beyond the Standard Model (SM). As reviewed below, QCD permits the existence of a CP- and P-violating term involving the contraction of the gluon field strength tensor with its dual. The coefficient of that term, $\bar{\theta}$, is however strongly constrained by the failure of highly-sensitive experiments to measure a nonzero neutron electric dipole moment, with the current bound being $|\bar{\theta}| \lesssim 3.5 \times 10^{-11}$ \cite{Abel:2020pzs}. The strong CP problem is the mystery of why this parameter is so small in reality, despite theory permitting it to be of order one.

A leading general solution to the problem requires extending the SM to include a spontaneously broken global $U(1)$ symmetry that has a colour anomaly at the quantum level: the Peccei-Quinn (PQ) symmetry~\cite{Peccei:1977hh, Peccei:1977ur}. In this approach, the coefficient of the $\bar{\theta}$-term becomes a dynamical pseudoscalar field whose potential has a global minimum at $\bar{\theta} = 0$. As a famous and welcome bonus, the associated light pseudo-Nambu-Goldstone boson -- the axion~\cite{Weinberg:1977ma,Wilczek:1977pj} -- is a plausible dark matter candidate in certain regions of the axion's parameter space. 

Peccei-Quinn symmetry is a general framework, not a specific theory. Within that framework, there are two leading implementations, known as Kim-Shifman-Vainshtein-Zakharov (KSVZ)~\cite{Kim:1979if, Shifman:1979if}  and Dine-Fischler-Srednicki-Zhitnitskii (DFSZ)~\cite{Dine:1981rt, Zhitnitsky:1980tq} models. In both cases, the axion is ``invisible'', meaning that its couplings to SM particles are suppressed and thus phenomenologically acceptable. This is achieved by introducing a complex scalar field, the PQ scalar, that is charged under $U(1)_{\text{PQ}}$ but is a singlet of the SM gauge group, so it has no Yukawa couplings to SM fermions. The modulus of this scalar is required to develop a large vacuum expectation value (VEV), while the phase is the dominant component of the axion field admixture. 

The difference between the two approaches lies in how the colour anomaly is generated. In KSVZ models an exotic quark that Yukawa couples to the PQ scalar is introduced, while in the DFSZ models a second electroweak Higgs doublet is added to the SM particle content. The latter models thus have the potential to produce observable high-energy collider signatures.

The present paper is motivated by a well-known cosmological problem of the standard DFSZ model~\cite{Sikivie:1982qv}: the existence of stable domain walls (DWs) whose energy density could unacceptably dominate the universe~\cite{Zeldovich:1974uw}. As reviewed in the next section, this problem arises because QCD instanton effects explicitly break $U(1)_{\text{PQ}}$ to a non-trivial exact discrete subgroup, so the theory has discrete degenerate vacua. Spontaneous breaking of the discrete symmetry then implies the existence of topologically stable DW solutions that interpolate between pairs of the discrete degenerate vacua. Cosmologically, these DWs can populate the universe via the Kibble mechanism.

Of the proposed solutions to this problem -- inflating away the DWs \cite{Guth:1980zm}, explicit PQ-breaking~\cite{Sikivie:1982qv, Barr:2014vva,Reig:2019vqh, Caputo:2019wsd}, embedding the PQ symmetry in a gauge~\cite{Lazarides:1982tw} or global group~\cite{Barr:1982bb} -- one is of particular interest because of its elegance and connection with the flavour puzzle of the SM. This solution sees the PQ symmetry become flavour-dependent in such a way that instanton effects explicitly break it \textit{completely}, so there is no exact discrete subgroup and thus no topologically stable DWs.

This approach was proposed in the 1980s, and some detailed analyses were undertaken~\cite{Davidson:1983tp,Davidson:1984ik,Geng:1988nc,Geng:1992jg}, but no general classification of three-family DW-free DFSZ models exists in the literature. The purpose of this paper is to provide such a classification. We shall see that specific cases that have already been analysed, such as the top-specific model \cite{Chiang:2017fjr, Chiang:2015cba, Saikawa:2019lng}, are instances within a rich catalogue of flavour-dependent, DW-free DFSZ models. Many of these new models warrant further investigation, including through the provision of a complete cosmological history, as was partially completed in Ref.~\cite{Sopov:2022bog} for the top-specific case.

The remainder of this paper is structured as follows. Section~\ref{sec:review} reviews the standard DFSZ model and the resulting cosmological domain wall problem, and then introduces flavour-dependent PQ charges as a solution strategy. The classification of three-family DFSZ models that have no DW problem is then provided in Sec.~\ref{sec:flavourvariant}. In Sec.~\ref{sec:texturezeros} we show that the PQ-induced texture zeros in the quark mass matrices do not imply any restrictions on the quark masses and CKM parameters, except in one model class. Sec.~\ref{sec:pheno} is a brief discussion of some of the phenomenological consequences of these models. We then conclude in Sec.~\ref{sec:conclusion}.

\newpage
\section{Review of the DFSZ model and the domain wall problem}
\label{sec:review}

The strong CP problem arises from the following term in the QCD Lagrangian:
\begin{equation}
    \mathcal{L}_{\theta}=\theta\frac{g_s^2}{32\pi^2}G_{\mu\nu}^a\tilde{G}^{a\mu\nu},
\end{equation}
where $G^a_{\mu \nu}$ is the gluon field strength tensor, $\tilde{G}^{a\mu\nu}=\frac{1}{2}\epsilon^{\mu\nu\alpha\beta}G_{\alpha\beta}^a$ is its dual and we use the convention that $\epsilon^{0123}=-1$.  As stated above, the PQ solution introduces a pseudoscalar field in such a way that the coefficient $\theta$ of the above term becomes dynamical, with the potential energy minimum occurring when the coefficient is zero, thus eliminating the strong CP problem. (We review the relation between $\theta$ and $\bar{\theta}$ below.)

In DFSZ implementations, the scalar field content comprises two Higgs doublets $\Phi_{1,2} \sim (1,2,1)$ and a complex singlet PQ scalar $S \sim (1,1,0)$, with the transformation properties under the SM gauge group $SU(3)_c \times SU(2)_L \times U(1)_Y$ as indicated. The PQ symmetry is defined to act on the quark and scalar fields as
\begin{align}
    q_{Li} &\rightarrow e^{i\mX(q_i)\alpha}q_{Li},\quad  u_{Ri} \rightarrow e^{-i\mX(u_i)\alpha} u_{Ri},\quad  d_{Ri} \rightarrow e^{-i\mX(d_i)\alpha}d_{Ri},\nonumber \\
    \Phi_r &\rightarrow e^{i\mX_r\alpha}\Phi_{r},\quad  S \rightarrow e^{i\mX_S \alpha}S,
    \label{pqsymm}
\end{align}
where $r \in \{1,2\}$ labels the Higgs doublets, $i$ is a family index and the $\mX$ are the PQ charges of the various fields. The special case of the standard DFSZ model has the PQ charges being family-independent, i.e.
\begin{equation}
\mX(q_i)=\mX(q),\ \ \mX(u_i)=\mX(u),\ \ \text{and}\ \ \mX(d_i)=\mX(d),
\end{equation}
for all $i$.

The gauge symmetry structure prevents $S$ from Yukawa-coupling to SM fermions, which ultimately makes the axion invisible. But $S$ must not completely decouple from the SM sector. If it did, there would actually be two new $U(1)$ symmetries: a PQ symmetry acting on just the quarks and the two Higgs doublets, and a second $U(1)$ that only transforms $S$. This would produce a visible axion and is thus ruled out. This circumstance is easily and naturally obviated by coupling $S$ to the Higgs doublets in a way that relates their PQ charges. There are two ways to do this, which correspond to the scalar potentials
\begin{equation}\label{potential}
    \begin{split}
        V(\Phi_1,\Phi_2,S) &= V_{\text{moduli}}+ \begin{cases} \lambda_1 \Phi_2^\dagger\Phi_1 S\ + \text{h.c.}\ \ \text{if}\ \mathcal{X}_2 - \mathcal{X}_1 =1 \hspace{5mm}(\text{Cubic Model}),\\
        \lambda_2 \Phi_2^\dagger\Phi_1 S^2 + \text{h.c.}\ \ \text{if}\ \mathcal{X}_2 - \mathcal{X}_1 = 2 \hspace{5mm}(\text{Quartic Model}),
        \end{cases}
    \end{split}
\end{equation}
where we have set $\mathcal{X}_S=1$ by convention, and $V_{\text{moduli}}$ is a function of $\Phi_1^\dagger \Phi_1$, $\Phi_2^\dagger \Phi_2$ and $S^* S$. As we will see, the cubic and quartic models give rise to different domain wall numbers $\mN_{DW}$, with drastic implications for cosmology.

To identify the axion field, we note that all three scalar multiplets contain fields with nonzero VEVs and adopt a convenient parameterisation given by
\begin{align} \label{eq:scalar-param}
    \Phi_1 &\supset \frac{v_1}{\sqrt{2}}e^{i\frac{a_{1}}{v_1}}\begin{pmatrix}
        0\\
        1
    \end{pmatrix}, & \Phi_2 &\supset \frac{v_2}{\sqrt{2}}e^{i\frac{a_{2}}{v_2}}\begin{pmatrix}
        0\\
        1
    \end{pmatrix}, &
     S &\supset \frac{v_{S}}{\sqrt{2}}e^{i\frac{a_{S}}{v_{S}}},
\end{align}
where $v_{\text{SM}} = \sqrt{v_1^2+v_2^2} \approx 246 \text{ GeV}$ and $v_S \gg v_1, v_2$. In this parameterisation, the charged modes and the neutral modulus modes have been neglected, while the neutral phase modes have been retained as the fields $a_{1,2,S}$. The physical axion, $a$, is then given by~\cite{Srednicki:1985xd}
\begin{align}
    a = \frac{1}{v_a}\sum_{i=1,2,S}\mathcal{X}_iv_ia_i\quad \text{with}\quad v_a^2=\sum_{i=1,2,S} \mathcal{X}_i^2v_i^2. 
    \label{axion}
\end{align}
Due to the condition $v_S \gg v_{\text{SM}}$, the axion field is primarily composed of $a_S$. This decoupling of the PQ scale from the electroweak scale leads to the suppression of the axion's mass and its couplings to SM fields.

The Yukawa Lagrangian must respect the PQ symmetry (\ref{pqsymm}) and, to be consistent with (\ref{potential}), must not enforce $\mathcal{X}_1 = \mathcal{X}_2$. Therefore, $\Phi_1$ and $\Phi_2$ cannot both couple to the same fermion bilinear; this is a necessary but not sufficient condition for $\mathcal{X}_1 \neq \mathcal{X}_2$. The Yukawa Lagrangian for the standard DFSZ model (with family-independent PQ charges) must therefore take the form,\footnote{Permuting $\Phi_1 \leftrightarrow \Phi_2$ would also yield a valid Yukawa Lagrangian. Also note that $\tilde{\Phi}_1=i\sigma_2\Phi_1^*$.}
\begin{equation} \label{eq:DFSZ-Yukawa}
    -\mathcal{L}_{Y}^{DFSZ} = Y^u\bar{q}_L\tilde{\Phi}_1u_R + Y^d\bar{q}_L\Phi_2d_R + \text{h.c.},
\end{equation}
where $Y^{u,d}$ are $3 \times 3$ complex matrices and we have suppressed the family indices. For this Lagrangian to respect the PQ symmetry, the relations
\begin{equation}
    \mX(q)+\mX(u) = - \mathcal{X}_1, \qquad \mX(q)+\mX(d) = \mathcal{X}_2,
    \label{pqrel}
\end{equation}
must hold between the PQ charges of the fermions and the Higgs doublets. 

The PQ symmetry is anomalous with respect to QCD and under the $U(1)_{PQ}$ transformation (\ref{pqsymm}) the coefficient of the $G\tilde{G}$ term shifts as $\theta \rightarrow \theta +2N\alpha$. The anomaly coefficient is in general given by
\begin{equation}
    2N = \sum_{i=1}^{N_f}\left[2\mX(q_i)+\mX(u_i)+\mX(d_i)\right] , \label{eq:PQ-anomaly} 
\end{equation}
where $N_f$ is the number of families. For the special case of family-independent PQ charges this evaluates to
\begin{equation}
      2N = N_f\left[2\mX(q)+\mX(u)+\mX(d)\right] = N_f (\mX_2 - \mX_1) , \label{eq:PQ-anomaly-DFSZ}
\end{equation}
where we have used the relations in (\ref{pqrel}). Given the $2\pi$ periodicity of $\theta$, it is clear that a discrete subgroup $\mathbb{Z}_{2N}$ of $U(1)_{PQ}$ is a symmetry of the $\theta$-term. The upshot is that QCD instanton effects explicitly break $U(1)_\text{PQ}$ to $\mathbb{Z}_{2N}$~\cite{Sikivie:1982qv}. 

This exact discrete symmetry of the path integral is, however, spontaneously broken, and it is well-known that this leads to domain walls~\cite{Zeldovich:1974uw}. The domain wall number is $\mN_{DW}=|2N|$, which is the number of inequivalent, degenerate vacua of the axion field. In the standard DFSZ model it reduces to $\mN_{DW}=N_f$ or $\mN_{DW}=2N_f$ for the case of a cubic or quartic potential, respectively~\cite{Geng:1990dv}.

To see how the domain walls arise in more detail, we begin by substituting \eqref{eq:scalar-param} into the Yukawa Lagrangian \eqref{eq:DFSZ-Yukawa} to obtain
\begin{equation} \label{eq:masslag}
    -\mathcal{L}_{mass}^{DFSZ} \supset \bar{u}_L Y^u u_R \frac{v_1}{\sqrt{2}} e^{-i\mathcal{X}_1 \frac{a}{v_a}} + \bar{d}_L Y^d d_R \frac{v_2}{\sqrt{2}} e^{i\mathcal{X}_2 \frac{a}{v_a}} + \text{h.c.} ,
\end{equation}
where we have used that $a_i/v_i = \mathcal{X}_i a/v_a + \ldots$, obtained by inverting \eqref{axion} and neglecting the other pseudoscalar modes. The physical strong CP angle, invariant under chiral quark rephasings, is $\bar\theta = \theta + \arg\det Y^u Y^d$, and it is useful to work in a basis where the coefficient of the $G\tilde{G}$ term in the Lagrangian is $\bar\theta$. Next, we perform a field redefinition under which the fermions (only) transform as in \eqref{pqsymm} with $\alpha=a/v_a$. This removes the axion from \eqref{eq:masslag} and moves it to the $G\tilde{G}$ term,
\begin{equation} \label{eq:aGG}
    \mathcal{L}_{\theta} \to \left( \bar{\theta}+\frac{2Na}{v_a} \right) \frac{g_s^2}{32\pi^2} G^a_{\mu\nu} \tilde{G}_a^{\mu\nu} ,
\end{equation}
where $f_a=v_a/(2N)$ is identified as the axion decay constant. This term gives rise to an instanton-generated potential for the axion in the low-energy effective theory below $\Lambda_{QCD}$~\cite{DiVecchia:1980yfw,GrillidiCortona:2015jxo},\footnote{We approximate the axion potential as a cosine for notational simplicity, but the discussion only relies on the fact that the potential is $2\pi$-periodic in $a\mN_{DW}/v_a$.}
\begin{equation}
    \label{poten}
    V(\varphi) \sim m_a^2f_a^2\cos \left( \frac{2Na}{v_a} + \bar{\theta}\right).
\end{equation}
The PQ symmetry acts as a shift on $a$, and the action of the discrete subgroup, 
\begin{equation}
    \mathbb{Z}_{\mN_{DW}}=\{\exp(2\pi i k/\mN_{DW})\,\,|\,\, k=0,1,...,\mN_{DW}-1\}, 
\end{equation}
with $\mN_{DW}=|2N|$, shifts $a \rightarrow a + 2\pi k v_a/\mN_{DW}$. This is a symmetry of the potential (\ref{poten}) that gets spontaneously broken when $a$ acquires a nonzero VEV. Minimising the potential yields the discrete, degenerate minima $\langle a \rangle_k= (-\bar{\theta}+2\pi k)v_a/\mN_{DW}$ for $k=0,1,..., \mN_{DW}-1$. The strong CP problem has been solved via the PQ mechanism, with the axion VEV dynamically cancelling $\bar\theta$ in \cref{eq:aGG}.

The degeneracy of the axion vacua, $\langle a \rangle_k$, gives rise to axionic domain walls in the early universe. This is due to the Kibble mechanism \cite{Kibble:1976sj}: when the universe cools to temperatures less than $\Lambda_{\text{QCD}}$, patches of the universe that are causally disconnected will settle into different vacua drawn from the set $\langle a \rangle_k$. The boundaries between these patches contain domain walls that smoothly interpolate between the neighbouring vacua and which contribute to the energy density of the universe. 

For $\mN_{DW} > 1$, these axionic domain walls are topologically stable and quickly dominate the energy density of the universe if PQ symmetry breaking happens after inflation~\cite{Marsh:2015xka, Sikivie:2006ni, Sikivie:1982qv}. This is known as the domain wall problem. The standard DFSZ model with either the cubic potential ($\mN_{DW} = 3$), or quartic potential ($\mN_{DW} = 6$) suffers from the domain wall problem.

One solution is to modify the DFSZ model such that $\mN_{DW} = 1$. In this case, domain walls still form in the early universe but their existence is short-lived. They rapidly become unstable and efficiently decay into non-relativistic axions that can make up a fraction of cold dark matter \cite{Chang:1998tb, Vilenkin:1982ks, Shellard:1986in}. While there are other solutions to the domain wall problem, as mentioned in the Introduction, we consider the $\mN_{DW} = 1$ solution, implemented through flavour-dependent PQ symmetries, to be elegant, robust and simple.

\section{Classification of flavour-variant models}
\label{sec:flavourvariant}

To construct $\mN_{DW}=1$ variants of the DFSZ model, we consider a family-dependent PQ symmetry as defined in \cref{pqsymm}. This modifies the colour anomaly with respect to the standard DFSZ model, and hence changes the domain wall number, without upsetting the PQ mechanism.

Flavour-dependent PQ symmetries are intriguing, since they have the potential to shed light on the enigmatic flavour structure of the SM. They were first considered by Bardeen and Tye~\cite{Bardeen:1977bd}, followed later by Davidson and Vozmediano \cite{Davidson:1983tp,Davidson:1984ik}. The latter work identified the connection between the vacuum and flavour structures and constructed all flavoured PQ symmetries with $\mN_{DW}=1$ for $N_f=2$. Flavoured PQ symmetries have been further motivated via a connection with the horizontal flavour symmetries of Grand Unified Theories \cite{Davidson:1981zd,Davidson:1983fy} or the Frogatt-Nielson mechanism \cite{Calibbi:2016hwq, Ema:2018abj, Arias-Aragon:2017eww}.

While some specific flavoured PQ symmetries with unit domain wall number have been investigated in the literature \cite{Peccei:1986pn,Geng:1988nc,Geng:1992jg,Chen:2010su,Chiang:2015cba,Chiang:2017fjr,Sun:2020iim,Sopov:2022bog} (for KSVZ-type models see~\cite{Alonso-Alvarez:2023wig}), in this paper we provide the first exhaustive catalogue encompassing all three-family, unit domain wall number, flavoured DFSZ models.

In order to construct such a catalogue, we first discuss how the anomaly coefficient, $2N$, is modified in the case of a flavoured PQ symmetry. In the previous section, we saw that the PQ relations (\ref{pqrel}) could be used to express the colour anomaly in terms of the Higgs PQ charges (\ref{eq:PQ-anomaly-DFSZ}). For the case of a family-dependent PQ symmetry, a similar relation can be derived. Consider a general Yukawa Lagrangian invariant under a flavour-dependent PQ symmetry. It contains terms of the form
\begin{align}
    \bar{q}_{Li}\tilde{\Phi}_ru_{Rj}, \text{\quad \quad and \quad \quad} \bar{q}_{Lk}\Phi_sd_{R\ell},
    \label{billinears}
\end{align}
where $r,s \in \{1,2\}$ and $i,j,k,\ell \in \{1,2,...,N_f\}$. These terms imply the following relations between the PQ charges:
\begin{align}
    \mX(q_i)+\mX(u_j)=-\mathcal{X}_r \text{\quad \quad and \quad \quad} \mX(q_k)+\mX(d_{\ell})=\mathcal{X}_s.
    \label{inducedcharges}
\end{align}
Let us first focus on the up-quark sector. We demand that the determinant of the mass matrix is non-zero so that none of the quarks are massless. This implies that there exists at least one subset of $N_f$ operators in the Yukawa Lagrangian where each left- and right-handed fermion appears only once (e.g., all the diagonal operators). Each of these operators gives rise to a PQ charge relation of the form (\ref{inducedcharges}). Let $n_u$ be the number of operators in this $N_f$ operator subset that couple to $\Phi_2$. This implies that there are $N_f-n_u$ operators in this subset which couple to $\Phi_1$, since $\Phi_1$ and $\Phi_2$ cannot couple to the same bilinear. Then, summing the PQ relations arising from these $N_f$ operators gives
\begin{equation}
    \label{upcha}
    \sum_i \mX(q_i)+\sum_j \mX(u_j) = (N_f-n_u)(-\mathcal{X}_1)+n_u(-\mathcal{X}_2).
\end{equation}
An analogous analysis in the down sector yields
\begin{equation}
    \label{downcha}
    \sum_i \mX(q_i) + \sum_j \mX(d_j) = (N_f-n_d)\mathcal{X}_1 +n_d \mathcal{X}_2.
\end{equation}
Summing (\ref{upcha}) and (\ref{downcha}) gives the overall colour anomaly in terms of the charges of the Higgs doublets:
\begin{equation}
   2N = \sum_{i=1}^{N_f}\left[2\mX(q_i)+\mX(u_i)+\mX(d_i)\right]=(n_d-n_u)(\mathcal{X}_2-\mathcal{X}_1)= n(\mathcal{X}_2-\mathcal{X}_1) \,,
    \label{colour}
\end{equation}
where $n\equiv n_d-n_u$ satisfies $-N_f \le n\le N_f$ and generalises the $n=N_f$ of the standard DFSZ model. As before, $\theta \rightarrow \theta + 2N\alpha$ under a PQ transformation \eqref{pqsymm}, and there exists a spontaneously broken $\mathbb{Z}_{2N}$ discrete symmetry. This leads to domain wall numbers $\mN_{DW}=|n|$ and $\mN_{DW}=2|n|$ for the cubic and quartic scalar potentials, respectively. An immediate consequence of this is that the quartic models have even domain wall number and only the cubic scalar potential can be used to construct $\mN_{DW}=1$ models.\footnote{The $n=0$ case is disallowed because then the PQ mechanism fails.}

All $\mN_{DW}=1$ models must therefore satisfy the colour anomaly relation
\begin{equation}
    \mN_{DW}=|2N|=\bigg| \sum_{i=1}^{N_f}\left[2\mX(q_i)+\mX(u_i)+\mX(d_i)\right] \bigg| = |\mathcal{X}_2-\mathcal{X}_1|=1.
    \label{dm1}
\end{equation}
The above equation is invariant under any of the following interchanges
\begin{align}\label{symm}
    \mX(q_k) &\leftrightarrow \mX(q_{\ell}), & \mX(u_k) &\leftrightarrow \mX(u_{\ell}), & \mX(d_k) &\leftrightarrow \mX(d_{\ell}), & \mathcal{X}_1 &\leftrightarrow \mathcal{X}_2, & u &\leftrightarrow d,
\end{align}
for $k, \ell \in \{1,2,\ldots,N_f \}$. Hence, any Yukawa Lagrangian that gives rise to a colour anomaly relation that satisfies (\ref{dm1}) can be transformed using any of the above permutations to give another $\mN_{DW}=1$ model.

To construct the $\mN_{DW}=1$ Lagrangians for $N_f=3$, we introduce the following notation for the Yukawa Lagrangian, following Ref.~\cite{Davidson:1984ik}:
\begin{align}\label{notation}
    {\scriptstyle \bar{q}_{Li}}\Bigg\{\overbrace{\begin{pmatrix}
    \times & \times & \times \\
    \times & \times & \times \\
    \times & \times & \times \\
    \end{pmatrix}_{\!\!\!u}}^{u_{Rj}}, \quad \quad \quad \quad
        {\scriptstyle \bar{q}_{Li}}\Bigg\{\overbrace{\begin{pmatrix}
    \times & \times & \times \\
    \times & \times & \times \\
    \times & \times & \times \\
    \end{pmatrix}_{\!\!\!d}}^{d_{Rj}}. 
\end{align}
These ``Yukawa patterns'' represent the couplings of $\Phi_1$ and $\Phi_2$ in the up and down sectors, respectively. The $\times$ are placeholders for either $1,2$ or $\bullet$ which indicate a coupling to $\Phi_1$, $\Phi_2$ or a disallowed coupling to both, respectively. Each of the non-zero couplings then implies a PQ charge relation like that of \cref{inducedcharges}. Note that we assume that all couplings allowed by the PQ symmetry are non-zero; the absence of a coupling therefore implies a PQ charge inequality. In this notation, the exchanges in \cref{symm} correspond to permuting the rows or columns, interchanging $1 \leftrightarrow 2$, and interchanging the up and down patterns.\footnote{For $\mX(q_k) \leftrightarrow \mX(q_{\ell})$, the rows of the up and down sector must be permuted simultaneously.} 

To illustrate this notation, consider the following example for the down sector
\begin{align}
    \begin{pmatrix}
    1 & \bullet & \bullet \\
    1 & 2 & 2 \\
    \bullet & \bullet & 2
    \end{pmatrix}_{\!\!\!d} &\Leftrightarrow 
    \setlength\arraycolsep{2pt}
    \begin{Bmatrix}
    \mX(q_1)+\mX(d_1) &=& \mathcal{X}_1; & \mX(q_1)+\mX(d_2) &\neq& \mathcal{X}_r; & \mX(q_1)+\mX(d_3) &\neq& \mathcal{X}_r\\
    \mX(q_2)+\mX(d_1) &=& \mathcal{X}_1; & \mX(q_2)+\mX(d_2) &=&\mathcal{X}_2; & \mX(q_2)+\mX(d_3) &=& \mathcal{X}_2 \\
    \mX(q_3)+\mX(d_1) &\neq& \mathcal{X}_r; & \mX(q_3)+\mX(d_2) &\neq& \mathcal{X}_r; & \mX(q_3)+\mX(d_3) &=&\mathcal{X}_2 
    \end{Bmatrix} 
\end{align}
The above pattern corresponds to the following down sector Yukawa Lagrangian
\begin{align} \label{eq:YukawaExample}
    -\mathcal{L}_Y&=Y^d_{11}\bar{q}_{L1}\Phi_1d_{R1}+ Y^d_{21}\bar{q}_{L2}\Phi_1d_{R1}+Y_{22}^d\bar{q}_{L2}\Phi_2d_{R2}
    +Y^d_{23}\bar{q}_{L2}\Phi_2d_{R3}+Y^d_{33}\bar{q}_{L3}\Phi_2d_{R3}+\text{h.c.},
\end{align}
where $Y^d_{ij}$ are the Yukawa couplings. This flavour variant may be contrasted with the standard DFSZ model which has
\begin{align}
    \begin{pmatrix}
    1 & 1 & 1 \\
    1 & 1 & 1 \\ 
    1 & 1 & 1
    \end{pmatrix}_{\!\!\!u} \oplus \begin{pmatrix}
    2 & 2 & 2 \\
    2 & 2 & 2 \\ 
    2 & 2 & 2
    \end{pmatrix}_{\!\!\!d}
\end{align}
as its Yukawa pattern.

To construct the flavoured $\mN_{DW}=1$ variant models we begin by fixing the diagonal entries of the Yukawa patterns in both the up and down sectors. As discussed previously, this is sufficient to determine the colour anomaly and hence the domain wall number. There exist three inequivalent diagonal patterns that give rise to $\mN_{DW}=1$, which we present in Table~\ref{tab:diagonal}. It can be readily verified that each of these yield $\mN_{DW}=1$ by summing the corresponding charge relations. All other possible diagonal couplings which fix $\mN_{DW}=  1$ are related to one of the diagonals in Table~\ref{tab:diagonal} through the permutations in \cref{symm}. 
\begin{table}[t]
    \centering
    \begin{tabular}{ccc}\toprule 
    Diagonal & Pattern & Anomaly Coefficient \\ \midrule
    \cellcolor{green!20}$\mathbf{D_1}$ &\cellcolor{green!20}$\begin{pmatrix}
    1 & & \\
    & 1 & \\
    & & 2
    \end{pmatrix}_{\!\!\!u} \oplus \begin{pmatrix}
    1 & & \\
    & 1 & \\
    & & 1
    \end{pmatrix}_{\!\!\!d}$&\cellcolor{green!20} \hspace{2.5mm}$2N=-1$\\ [0.6cm]
   \cellcolor{blue!20}$\mathbf{D_2}$&\cellcolor{blue!20}$\begin{pmatrix}
    1 & & \\
    & 1 & \\
    & & 2
    \end{pmatrix}_{\!\!\!u} \oplus\begin{pmatrix}
    2 & & \\
    & 2 & \\
    & & 1
    \end{pmatrix}_{\!\!\!d} $ & \cellcolor{blue!20}$2N=1$\\[0.6cm] 
    \cellcolor{pink!20}$\mathbf{D_3}$&\cellcolor{pink!20}$\begin{pmatrix}
    1 & & \\
    & 1 & \\
    & & 2
    \end{pmatrix}_{\!\!\!u} \oplus \begin{pmatrix}
    1 & & \\
    & 2 & \\
    & & 2
    \end{pmatrix}_{\!\!\!d}$ &\cellcolor{pink!20} $2N=1$ \\[0.6cm] 
    \bottomrule
    \end{tabular}
    \caption{The three pairs of diagonal Yukawa coupling patterns that fix $\mN_{DW}=1$ and which are inequivalent under the permutations (\ref{symm}). Off-diagonal couplings are yet to be specified.} 
\label{tab:diagonal}
\end{table}

The off-diagonal entries of the Yukawa patterns in Table~\ref{tab:diagonal} must now be filled with $1$, $2$ or $\bullet$. This must be done so that the Yukawa Lagrangians are fully specified in the sense that all bilinears of the Yukawa Lagrangian are either coupled to $\Phi_1$ or $\Phi_2$, or both couplings are disallowed by the PQ symmetry. Consider a given diagonal up and down sector $\mathbf{D_i}$ in Table~\ref{tab:diagonal}. In principle, there are $3^{12}$ different ``patterns'' or ways to specify the off-diagonal couplings; however, the  viable $\mN_{DW}=1$ models are the subset of these $3^{12}$ patterns which comply with the following conditions:
\begin{enumerate}
\item $\mX_1 \neq \mX_2$. This condition is required for consistency with the cubic potential term (\ref{potential}) and to guarantee that the PQ colour anomaly coefficient, $2N$,  in \cref{colour} is non-zero. Combinations of off-diagonal couplings can lead to $\mX_1 =\mX_2$ as a result of the charge relations (\ref{inducedcharges}). 
\item $U(1)_{\text{PQ}} \times U(1)_{\text{B}}$ are the only global $U(1)$ symmetries respected by the Yukawa Lagrangian. All operators which are invariant under this symmetry are included in the Yukawa Lagrangian.\footnote{The viable $\mN_{DW}=1$ models may contain ``subset models'' that have additional $U(1)$ symmetries in the limit where some of the coupling constants go to zero. These subset models are technically natural, but may be too constrained to reproduce the measured mixing angles of the quark sector.}
\end{enumerate}
Iterating over all possible off-diagonal patterns (for each diagonal $\mathbf{D_i}$) and removing those that do not satisfy the above conditions yields all viable $\mN_{DW}=1$ models. We group these models into equivalence classes under the permutations in \cref{symm}. A single representative model for each class is listed in \cref{tab:fulllist}.

\begin{table}[t]
    \centering
    \begin{tabular}{p{1cm} c p{0.5cm} p{1cm} c}\toprule 
    Class & Yukawa Pattern & & Class & Yukawa Pattern \\ \midrule
    \cellcolor{green!20}$\mathbf{D_{1,1}}$ & \cellcolor{green!20}$\begin{pmatrix}
    1 & 1 & 2 \\
    1 & 1 & 2 \\ 
    1 & 1 & 2
    \end{pmatrix}_{\!\!\!u} \oplus \begin{pmatrix}
    1 & 1 & 1 \\
    1 & 1 & 1 \\ 
    1 & 1 & 1
    \end{pmatrix}_{\!\!\!d}$  
    & \cellcolor{green!20} &\cellcolor{green!20}$\mathbf{D_{1,2}}$ &\cellcolor{green!20}$\begin{pmatrix}
    1 & \bullet & 1 \\
    2 & 1 & 2 \\ 
    2 & 1 & 2
    \end{pmatrix}_{\!\!\!u}\oplus \begin{pmatrix}
    1 & 2 & 2 \\
    \bullet & 1 & 1 \\ 
    \bullet & 1 & 1
    \end{pmatrix}_{\!\!\!d}$  \\[0.3cm] 
    \cellcolor{green!20}$\mathbf{D_{1,3}}$ &\cellcolor{green!20}$\begin{pmatrix}
    1 & 2 & \bullet \\
    \bullet & 1 & 2 \\ 
    \bullet & 1 & 2
    \end{pmatrix}_{\!\!\!u}\oplus\begin{pmatrix}
    1 & \bullet & \bullet \\
    2 & 1 & 1 \\ 
    2 & 1 & 1
    \end{pmatrix}_{\!\!\!d}$  
    & \cellcolor{green!20} &\cellcolor{green!20}$\mathbf{D_{1,4}}$ &\cellcolor{green!20}$\begin{pmatrix}
    1 & 1 & 1 \\
    1 & 1 & 1 \\ 
    2 & 2 & 2
    \end{pmatrix}_{\!\!\!u}\oplus\begin{pmatrix}
    1 & 1 & 2 \\
    1 & 1 & 2 \\ 
    \bullet & \bullet & 1
    \end{pmatrix}_{\!\!\!d}$ \\[0.3cm]
    \cellcolor{green!20}$\mathbf{D_{1,5}}$&\cellcolor{green!20}$\begin{pmatrix}
    1 & 1 & \bullet \\
    1 & 1 & \bullet \\ 
    \bullet & \bullet & 2
    \end{pmatrix}_{\!\!\!u}\oplus \begin{pmatrix}
    1 & 1 & \bullet \\
    1 & 1 & \bullet \\ 
    2 & 2 & 1
    \end{pmatrix}_{\!\!\!d}$    
    & \cellcolor{green!20} &\cellcolor{green!20}$\mathbf{D_{1,6}}$& \cellcolor{green!20}$\begin{pmatrix}
    1 & \bullet & \bullet \\
    2 & 1 & 1 \\ 
    \bullet & 2 & 2
    \end{pmatrix}_{\!\!\!u}\oplus \begin{pmatrix}
    1 & 2 & \bullet \\
    \bullet & 1 & 2 \\ 
    \bullet & \bullet & 1
    \end{pmatrix}_{\!\!\!d}$ \\[0.3cm] 
    \cellcolor{green!20}$\mathbf{D_{1,7}}$&\cellcolor{green!20}$\begin{pmatrix}
    1 & \bullet & \bullet \\
    \bullet & 1 & 1 \\ 
    \bullet & 2 & 2
    \end{pmatrix}_{\!\!\!u}\oplus \begin{pmatrix}
    1 & \bullet & \bullet \\
    \bullet & 1 & 2 \\ 
    2 & \bullet & 1
    \end{pmatrix}_{\!\!\!d}$ & \cellcolor{green!20} &\cellcolor{green!20}$\mathbf{D_{1,8}}$&\cellcolor{green!20}$\begin{pmatrix}
    1 & \bullet & \bullet \\
    2 & 1 & \bullet \\ 
    \bullet & \bullet & 2
    \end{pmatrix}_{\!\!\!u}\oplus \begin{pmatrix}
    1 & 2 & \bullet \\
    \bullet & 1 & \bullet \\ 
    2 & \bullet & 1
    \end{pmatrix}_{\!\!\!d}$  \\[0.6cm] 
    \cellcolor{blue!20}$\mathbf{D_{2,1}}$ & \cellcolor{blue!20}$\begin{pmatrix}
    1 & 1 & 2\\
    1 & 1 & 2\\
    1 & 1 & 2
    \end{pmatrix}_{\!\!\!u}\oplus \begin{pmatrix}
    2 & 2 & 1\\
    2 & 2 & 1\\
    2 & 2 & 1
    \end{pmatrix}_{\!\!\!d}$  
    & \cellcolor{blue!20} &\cellcolor{blue!20}$\mathbf{D_{2,2}}$ &\cellcolor{blue!20}$\begin{pmatrix}
    1 & \bullet & 1\\
    2 & 1 & 2\\
    2 & 1 & 2
    \end{pmatrix}_{\!\!\!u} \oplus \begin{pmatrix}
    2 & \bullet & 2\\
    1 & 2 & 1\\
    1 & 2 & 1
    \end{pmatrix}_{\!\!\!d}$ \\[0.3cm] 
    \cellcolor{blue!20}$\mathbf{D_{2,3}}$  &\cellcolor{blue!20}$\begin{pmatrix}
    1 & 2 & \bullet\\
    \bullet & 1 & 2\\
    \bullet & 1 & 2
    \end{pmatrix}_{\!\!\!u} \oplus \begin{pmatrix}
    2 & 1 & \bullet\\
    \bullet & 2 & 1\\
    \bullet & 2 & 1
    \end{pmatrix}_{\!\!\!d}$   
    & \cellcolor{blue!20} &\cellcolor{blue!20}$\mathbf{D_{2,4}}$ & \cellcolor{blue!20}$\begin{pmatrix}
    1 & 1 & 1\\
    1 & 1 & 1\\
    2 & 2 & 2
    \end{pmatrix}_{\!\!\!u}\oplus \begin{pmatrix}
    2 & 2 & 2\\
    2 & 2 & 2\\
    1 & 1 & 1
    \end{pmatrix}_{\!\!\!d}$ \\[0.3cm] 
    \cellcolor{blue!20}$\mathbf{D_{2,5}}$ & \cellcolor{blue!20}$\begin{pmatrix}
    1 & \bullet & \bullet\\
    2 & 1 & 1\\
    \bullet& 2 & 2
    \end{pmatrix}_{\!\!\!u}\oplus \begin{pmatrix}
    2 & \bullet & \bullet\\
    1 & 2 & 2\\
    \bullet & 1 & 1
    \end{pmatrix}_{\!\!\!d}$ & \cellcolor{blue!20} & \cellcolor{blue!20} & \cellcolor{blue!20} \\[0.6cm] 
    \cellcolor{pink!20}$\mathbf{D_{3,1}}$ &\cellcolor{pink!20}$\begin{pmatrix}
    1 & \bullet & 1\\
    2 & 1 & 2\\
    2& 1 & 2
    \end{pmatrix}_{\!\!\!u}\oplus \begin{pmatrix}
    1 & \bullet & \bullet\\
    \bullet & 2 & 2\\
    \bullet & 2 & 2
    \end{pmatrix}_{\!\!\!d}$  
    & \cellcolor{pink!20} &\cellcolor{pink!20} $\mathbf{D_{3,2}}$ &\cellcolor{pink!20}$\begin{pmatrix}
    1 & 2 & \bullet\\
    \bullet & 1 & 2\\
    \bullet& 1 & 2
    \end{pmatrix}_{\!\!\!u}\oplus \begin{pmatrix}
    1 & 1 & 1\\
    2 & 2 & 2\\
    2 & 2 & 2
    \end{pmatrix}_{\!\!\!d}$ \\[0.3cm] 
    \cellcolor{pink!20}$\mathbf{D_{3,3}}$ &\cellcolor{pink!20}$\begin{pmatrix}
    1 & \bullet & 1\\
    \bullet & 1 & \bullet\\
    2& \bullet & 2
    \end{pmatrix}_{\!\!\!u}\oplus \begin{pmatrix}
    1 & \bullet & \bullet\\
    \bullet & 2 & 1\\
    \bullet & \bullet & 2
    \end{pmatrix}_{\!\!\!d}$  
    & \cellcolor{pink!20} &\cellcolor{pink!20}$\mathbf{D_{3,4}}$ &\cellcolor{pink!20}$\begin{pmatrix}
    1 & \bullet & \bullet \\
    2 & 1 & 1 \\ 
    \bullet & 2 & 2
    \end{pmatrix}_{\!\!\!u}\oplus \begin{pmatrix}
    1 & \bullet & \bullet \\
    \bullet & 2 & \bullet\\ 
    \bullet & 1 & 2
    \end{pmatrix}_{\!\!\!d}$ \\[0.3cm] 
    \cellcolor{pink!20} $\mathbf{D_{3,5}}$ & \cellcolor{pink!20}$\begin{pmatrix}
    1 & \bullet & \bullet \\
    \bullet & 1 & 1 \\ 
    \bullet & 2 & 2
    \end{pmatrix}_{\!\!\!u}\oplus \begin{pmatrix}
    1 & \bullet & 1 \\
    \bullet & 2 & \bullet \\ 
    2 & 1 & 2
    \end{pmatrix}_{\!\!\!d}$ & \cellcolor{pink!20} & \cellcolor{pink!20} & \cellcolor{pink!20}\\
    \bottomrule
    \end{tabular}
    \caption{A representative model of each $\mN_{DW}=1$ equivalence class. Each class is labelled $\mathbf{D_{i,j}}$, where $\mathbf{i}=1,2,3$ labels the corresponding diagonal from \cref{tab:diagonal} and $\mathbf{j}$ enumerates the classes arising from that diagonal. The colours also indicate the corresponding diagonal pattern. The matrix entries $1,2$ and $\bullet$ denote a coupling to $\Phi_1$, $\Phi_2$ or a disallowed coupling to both, respectively. Note that of the eighteen classes, all except for $\mathbf{D_{3,3}}$ can reproduce the ten observables of the quark sector.}
    \label{tab:fulllist}
\end{table}

With the Yukawa patterns specified, the PQ charges for each model are also determined. We impose that the PQ current is orthogonal to the hypercharge current\footnote{There is freedom to redefine the PQ charges through the admixture of colour-anomaly-free weak hypercharge and baryon number.}, which enforces $\mX_1v_1^2+\mX_2v_2^2=0$ and guarantees there is no kinetic mixing between the axion and the $Z$ boson. This orthogonality condition, combined with the cubic potential term (\ref{potential}), fixes $\mX_1 = -\sin^2{\beta} \equiv -s_{\beta}^2$ and $\mX_2 = \cos^2{\beta} \equiv c_{\beta}^2$, where $\tan\beta \equiv v_2/v_1$. The remaining PQ charges can then be determined from the Yukawa matrix patterns, up to an admixture of the colour-anomaly-free baryon number. This freedom is used to set $\mX(q_3) = 0$ by convention. The PQ charges are the same (up to permutations) for all models within each equivalence class and are given in Table~\ref{PQcharges}.

\begin{table}[t]
    \centering
    \begin{tabular}{cccc}\toprule 
    Class & $\{\mX(q_1), \mX(q_2), \mX(q_3)\}$ & $\{\mX(u_1), \mX(u_2), \mX(u_3)\}$ & $\{\mX(d_1), \mX(d_2), \mX(d_3)\}$ \\ \midrule
   \cellcolor{green!20}$\mathbf{D_{1,1}}$ &\cellcolor{green!20}$\{0,0,0 \}$ & \cellcolor{green!20}$\{s_{\beta}^2,s_{\beta}^2,-c_{\beta}^2 \}$ & \cellcolor{green!20}$\{-s_{\beta}^2,-s_{\beta}^2,-s_{\beta}^2 \}$ \\ [0.1cm]
   \cellcolor{green!20}$\mathbf{D_{1,2}}$&\cellcolor{green!20}$\{1,0,0 \}$  & \cellcolor{green!20}$\{-c_{\beta}^2,s_{\beta}^2,-c_{\beta}^2 \}$ & \cellcolor{green!20}$\{-1-s_{\beta}^2,-s_{\beta}^2,-s_{\beta}^2 \}$ \\[0.1cm] 
    \cellcolor{green!20}$\mathbf{D_{1,3}}$&\cellcolor{green!20}$\{-1,0,0 \}$  &\cellcolor{green!20}$\{1+s_{\beta}^2,s_{\beta}^2,-c_{\beta}^2\}$  &\cellcolor{green!20}$\{c_{\beta}^2,-s_{\beta}^2,-s_{\beta}^2 \}$  \\[0.1cm] 
    \cellcolor{green!20}$\mathbf{D_{1,4}}$&\cellcolor{green!20}$\{1,1,0\}$  & \cellcolor{green!20}$\{-c_{\beta}^2,-c_{\beta}^2,-c_{\beta}^2 \}$ &\cellcolor{green!20}$\{-1-s_{\beta}^2,-1-s_{\beta}^2,-s_{\beta}^2\}$  \\ [0.1cm]
    \cellcolor{green!20}$\mathbf{D_{1,5}}$&\cellcolor{green!20}$\{-1,-1,0 \}$  &\cellcolor{green!20}$\{1+s_{\beta}^2,1+s_{\beta}^2,-c_{\beta}^2 \}$ &\cellcolor{green!20}$\{c_{\beta}^2,c_{\beta}^2,-s_{\beta}^2\}$\\[0.1cm] 
    \cellcolor{green!20}$\mathbf{D_{1,6}}$& \cellcolor{green!20}$\{2,1,0 \}$ &\cellcolor{green!20}$\{-1-c_{\beta}^2,-c_{\beta}^2,-c_{\beta}^2\}$  &\cellcolor{green!20}$\{-2-s_{\beta}^2,-1-s_{\beta}^2,-s_{\beta}^2 \}$ \\ [0.1cm]
    \cellcolor{green!20}$\mathbf{D_{1,7}}$&\cellcolor{green!20}$\{-1,1,0 \}$ & \cellcolor{green!20}$\{1+s_{\beta}^2,-c_{\beta}^2,-c_{\beta}^2 \}$ &\cellcolor{green!20}$\{c_{\beta}^2,-1-s_{\beta}^2,-s_{\beta}^2\}$\\[0.1cm] 
    \cellcolor{green!20}$\mathbf{D_{1,8}}$&\cellcolor{green!20}$\{-1,-2,0\}$ &\cellcolor{green!20}$\{1+s_{\beta}^2,2+s_{\beta}^2,-c_{\beta}^2 \}$  &\cellcolor{green!20}$\{c_{\beta}^2,1+c_{\beta}^2,-s_{\beta}^2\}$\\[0.1cm] 
    \cellcolor{blue!20}$\mathbf{D_{2,1}}$ &\cellcolor{blue!20}$\{0,0,0 \}$ &\cellcolor{blue!20}$\{s_{\beta}^2,s_{\beta}^2,-c_{\beta}^2 \}$ &\cellcolor{blue!20}$\{c_{\beta}^2,c_{\beta}^2,-s_{\beta}^2 \}$ \\[0.1cm] 
    \cellcolor{blue!20}$\mathbf{D_{2,2}}$ &\cellcolor{blue!20}$\{1,0,0 \}$  &\cellcolor{blue!20}$\{-c_{\beta}^2,s_{\beta}^2,-c_{\beta}^2 \}$ &\cellcolor{blue!20}$\{-s_{\beta}^2,c_{\beta}^2,-s_{\beta}^2 \}$\\[0.1cm] 
    \cellcolor{blue!20}$\mathbf{D_{2,3}}$ &\cellcolor{blue!20}$\{-1,0,0 \}$  &\cellcolor{blue!20}$\{1+s_{\beta}^2,s_{\beta}^2,-c_{\beta}^2 \}$ &\cellcolor{blue!20} $\{1+c_{\beta}^2,c_{\beta}^2,-s_{\beta}^2 \}$ \\ [0.1cm]
    \cellcolor{blue!20}$\mathbf{D_{2,4}}$ &\cellcolor{blue!20}$\{1,1,0\}$  & \cellcolor{blue!20}$\{-c_{\beta}^2,-c_{\beta}^2,-c_{\beta}^2 \}$ &\cellcolor{blue!20}$\{-s_{\beta}^2,-s_{\beta}^2,-s_{\beta}^2\}$\\[0.1cm] 
    \cellcolor{blue!20}$\mathbf{D_{2,5}}$ &\cellcolor{blue!20}$\{2,1,0 \}$ &\cellcolor{blue!20}$\{-1-c_{\beta}^2,-c_{\beta}^2,-c_{\beta}^2\}$  &\cellcolor{blue!20}$\{-1-s_{\beta}^2,-s_{\beta}^2,-s_{\beta}^2 \}$ \\[0.1cm] 
    \cellcolor{pink!20}$\mathbf{D_{3,1}}$ &\cellcolor{pink!20}$\{1,0,0 \}$ &\cellcolor{pink!20}$\{-c_{\beta}^2,s_{\beta}^2,-c_{\beta}^2\}$  &\cellcolor{pink!20}$\{-1-s_{\beta}^2,c_{\beta}^2,c_{\beta}^2 \}$ \\[0.1cm]
    \cellcolor{pink!20}$\mathbf{D_{3,2}}$ &\cellcolor{pink!20}$\{-1,0,0 \}$ &\cellcolor{pink!20}$\{1+s_{\beta}^2,s_{\beta}^2,-c_{\beta}^2\}$  &\cellcolor{pink!20} $\{c_{\beta}^2,c_{\beta}^2,c_{\beta}^2 \}$\\[0.1cm]
    \cellcolor{pink!20}$\mathbf{D_{3,3}}$ &\cellcolor{pink!20}$\{1,-1,0 \}$ &\cellcolor{pink!20}$\{-c_{\beta}^2,1+s_{\beta}^2,-c_{\beta}^2\}$  &\cellcolor{pink!20} $\{-1-s_{\beta}^2,1+c_{\beta}^2,c_{\beta}^2 \}$\\[0.1cm]
    \cellcolor{pink!20}$\mathbf{D_{3,4}}$ &\cellcolor{pink!20}$\{2,1,0 \}$ &\cellcolor{pink!20}$\{-1-c_{\beta}^2,-c_{\beta}^2,-c_{\beta}^2\}$  &\cellcolor{pink!20}$\{-2-s_{\beta}^2,-s_{\beta}^2,c_{\beta}^2 \}$\\[0.1cm]
    \cellcolor{pink!20}$\mathbf{D_{3,5}}$ &\cellcolor{pink!20}$\{-1,1,0 \}$ &\cellcolor{pink!20}$\{1+s_{\beta}^2,-c_{\beta}^2,-c_{\beta}^2\}$  & \cellcolor{pink!20}$\{c_{\beta}^2,-s_{\beta}^2,c_{\beta}^2 \}$\\[0.1cm]
    \bottomrule
    \end{tabular}
    \caption{\label{PQcharges}PQ charges for each of the representative $\mN_{DW}=1$ models in Table~\ref{tab:fulllist}. By convention $\mX(q_3)$ has been set to zero, and $\tan\beta \equiv v_2/v_1$. The colours indicate which diagonal the models arise from in Table~\ref{tab:diagonal}.}
\end{table}

Each equivalence class contains a large number of models, which are generated from the representative models in Table~\ref{tab:fulllist} via the permutations~\eqref{symm}. It should be emphasised that while the permutations leave $\mN_{DW}$ invariant, models belonging to the same equivalence class can have very different phenomenology. Since the PQ charges are non-universal, these models give rise to flavour-changing neutral processes mediated by the electroweak Higgs bosons and flavour-violating axion interactions, both of which are briefly discussed in \cref{sec:pheno}.

Some of the models contained in Table~\ref{tab:fulllist} have already been identified and studied in the literature. For example, the representative of class $\mathbf{D_{1,1}}$ is the top-specific model which, along with some of its permutations, was considered in Refs.~\cite{Chiang:2015cba,Chiang:2017fjr, Chen:2010su, Badziak:2021apn, Saikawa:2019lng, DiLuzio:2017ogq}. Models that belong to the equivalence classes represented by $\mathbf{D_{1,8}}$ and $\mathbf{D_{2,4}}$ were considered in Refs.~\cite{DiLuzio:2021ysg,Bjorkeroth:2018ipq, Badziak:2021apn, DiLuzio:2017ogq, Dias:2020kbj}. We note that the classes $\mathbf{D_{1,1}}$, $\mathbf{D_{2,1}}$ and $\mathbf{D_{2,4}}$ have Yukawa patterns which are SM-like in the sense that they do not have any texture-zeros. Interestingly, the flavour structure of class $\mathbf{D_{1,8}}$ gives rise to a fully horizontal PQ symmetry, distinguishing all left- and right-handed quark flavours.

We must still determine which models can accommodate the measured masses, mixing angles and CP-violating phase of the quark sector. The outcome is not \textit{a priori} obvious in all cases due to the existence of texture-zeros in the quark mass matrices. However, as shown in the next section, all classes of models in Table~\ref{tab:fulllist} are able to reproduce the physical SM parameters, with the exception of $\mathbf{D_{3,3}}$.

\section{Texture-zeros and the physical parameters of the quark sector}
\label{sec:texturezeros}

The texture-zeros in the Yukawa patterns in Table~\ref{tab:fulllist} can restrict the values of, or lead to relations between, the quark masses and/or CKM mixing angles and CP phase. While these relations could in principle provide a partial explanation for the observed SM flavour structure, they must not conflict with the measured values of the parameters in the quark sector. To determine whether our models give rise to such relations, it is convenient to rotate to the basis $(\Phi_1,\Phi_2) \to (H_1,H_2)$ where $\langle H_1 \rangle =\frac{v_{SM}}{\sqrt{2}}$ and $\langle H_2 \rangle=0$. The general Yukawa Lagrangian then takes the form
\begin{equation}\label{yukinhiggs}
    -\mathcal{L}_Y=\eta^u\bar{q}_{L}\tilde{H}_{1}u_{R}+\eta^d\bar{q}_{L}H_{1}d_{R}
    +\xi^u\bar{q}_{L}\tilde{H}_2u_{R}+\xi^d\bar{q}_{L}H_{2}d_{R} + \text{h.c.},
\end{equation}
where $\eta^q$ and $\xi^q$ are $3 \times 3$ complex Yukawa matrices defined by
\begin{align}\label{eq:higgsbasisyukawa}
    \eta^q &= c_{\beta}Y^q_{1}+s_{\beta}Y^q_{2}, \nonumber \\
    \xi^q &=-s_{\beta}Y^q_{1}+c_{\beta}Y^q_{2},
\end{align}
with $Y^q_{1}$ and $Y^q_{2}$ the couplings to the doublets $\Phi_1$ and $\Phi_2$, respectively, in the original basis. Note that this differs from the Yukawa notation used in \eqref{eq:YukawaExample}, where the subscripts denote the family index. The quark mass matrices $M_q=\frac{v_{SM}}{\sqrt{2}}\eta^q$ are then independent of $\xi^q$.

We first check whether the model classes in Table \ref{tab:fulllist} give rise to non-trivial masses and mixing angles. Since, by construction, $\det[M_q] \neq 0$ for all of the classes, none of them give rise to a massless quark. Now consider the quantity\footnote{This quantity is invariant under the row and column permutations in \eqref{symm} and transforms as $\mathcal{D} \to - \mathcal{D}$ under the interchange $u \to d$.} \cite{Jarlskog:1985ht,Jarlskog:1985cw}
\begin{align}
    \mathcal{D} = \det\left(\left[M_uM_u^{\dagger},M_dM_d^{\dagger}\right]\right) = 2iF_uF_d J,
\end{align}
where, in the standard parameterisation of the CKM matrix~\cite{ParticleDataGroup:2022pth},
\begin{align}
    F_u&=(m_t^2-m_c^2)(m_t^2-m_u^2)(m_c^2-m_u^2), \\
    F_d&=(m_b^2-m_s^2)(m_b^2-m_d^2)(m_s^2-m_d^2), \\ 
    J &= s_{12}c_{12}s_{23}c_{23}s_{13}c_{13}^2\sin\delta,
\end{align}
with $s_{ij} \equiv \sin\theta_{ij}$, $c_{ij}\equiv \cos\theta_{ij}$, and $J$ the Jarlskog invariant. The $\theta_{ij}$ are the three CKM mixing angles and $\delta$ is the CP-violating phase. It is clear that $\mathcal{D}$ can vanish only if there is a degeneracy of quark masses or any of the following conditions hold:
\begin{align}
   \theta_{ij} =0, \ \frac{\pi}{2}\quad \text{or}\quad \delta=0,\ \pi \,,
\end{align}
for any of the mixing angles $\theta_{ij}$. One can easily check that for class $\mathbf{D_{3,3}}$, with mass matrices
\begin{align}
    M_u=\begin{pmatrix}
        m_{11}^u&0&m_{13}^u \\
        0 & m_{22}^u&0 \\
        m_{31}^u&0&m_{33}^u
    \end{pmatrix}, \qquad     
    M_d=\begin{pmatrix}
        m_{11}^d&0&0 \\
        0 & m_{22}^d&m_{23}^d \\
       0&0&m_{33}^d
    \end{pmatrix},
\end{align}
the quantity $\mathcal{D}$ vanishes for arbitrary complex entries $m_{ij}^q$. All remaining classes yield a non-vanishing $\mathcal{D}$. This excludes class $\mathbf{D_{3,3}}$ from further consideration.

While all remaining classes have non-trivial masses and mixing angles, we still need to check whether there exist any relations between these parameters. To do this, we use the fact that SM-like mass matrices -- arbitrary $3 \times 3$ complex matrices with no texture-zeros -- do not induce any relations between the ten physical parameters. We then make use of weak basis transformations which can map an arbitrary matrix to one containing texture zeros, while preserving the masses and the CKM matrix~\cite{Branco:1988iq,Branco:1999nb}. A weak basis transformation is defined as
\begin{align} \boldsymbol{u}_L&= W^{\dagger}\boldsymbol{u}_L', & \boldsymbol{u}_R&= \tilde{V}_u\boldsymbol{u}_R', \nonumber \\
    \boldsymbol{d}_L&= W^{\dagger}\boldsymbol{d}_L', & \boldsymbol{d}_R&= \tilde{V}_d \boldsymbol{d}_R', 
    \label{wbt2}
\end{align}
or equivalently $M_{q}'=WM_{q}\tilde{V}_{q}$, where $W$, $\tilde{V}_u$ and $\tilde{V}_d$ are unitary matrices. The mass matrices $M_q'$ and $M_q$ yield the same quark masses and CKM parameters. Since permutation matrices are unitary, this immediately implies that transformations (\ref{symm}) between models of the same equivalence class also leave the masses and CKM observables invariant.

First, as an explicit example, we demonstrate that the mass matrices in class $\mathbf{D_{1,8}}$ are weak basis equivalent to SM-like mass matrices. As a convenient starting point, we perform a weak basis transformation to a basis where $M_u$ is a real diagonal matrix and $M_d$ is an arbitrary $3 \times 3$ complex matrix. Then, as illustrated in Ref.~\cite{Emmanuel-Costa:2016gdp}, we can utilize $\tilde{V}_d$ to introduce up to three texture-zeros in $M_d$. After applying these weak basis transformations, the mass matrices take the following form:
\begin{align}
    M_u = \begin{pmatrix}
        m_{11}^u&0&0\\
        0&m_{22}^u&0\\
        0&0&m_{33}^u
    \end{pmatrix}, \qquad
    M_d = \begin{pmatrix}
        m_{11}^d&m_{12}^d&0 \\
        m_{21}^d&m_{22}^d&0 \\
        m_{31}^d&0&m_{33}^d
    \end{pmatrix}.
\end{align}
By appropriate rephasing of the quark fields, we can render all elements of $M_u$ and $M_d$ real, except for $m_{22}^d$. Following this, we implement a weak basis transformation of the form
\begin{align}
    W=\begin{pmatrix}
        \cos\theta_1&-\sin\theta_1&0\\
        \sin\theta_1&\cos\theta_1&0\\
        0&0&1
    \end{pmatrix}, 
    \qquad
    \tilde{V}_u=\begin{pmatrix}
        \cos\theta_2&\sin\theta_2&0\\
        -\sin\theta_2&\cos\theta_2&0\\
        0&0&1
    \end{pmatrix},
    \qquad 
    \tilde{V}_d=\mathbb{1}.
\end{align}
Demanding that $m'^{d}_{21}=0$ and $m'^{u}_{12}=0$, where $m'^{q}_{ij}$ are the entries of $M'_q=WM_q\tilde{V}_q$, gives the relations
\begin{align}
    \tan\theta_1=-\frac{m_{21}^d}{m_{11}^d}, \qquad \tan\theta_2=\frac{m_{22}^u }{m_{11}^u} \tan\theta_1,
\end{align}
and leads to the texture pattern
\begin{align}
        M'_u = \begin{pmatrix}
        \times&0&0\\
        \times&\times&0\\
        0&0&\times
    \end{pmatrix}, \qquad
    M'_d = \begin{pmatrix}
        \times&\times&0 \\
        0&\times&0 \\
        \times&0&\times
    \end{pmatrix}.
    \label{mat:d18}
\end{align}
This is the texture-zero pattern of $\mathbf{D_{1,8}}$, thus demonstrating that this class of models does not impose any relations amongst the masses and CKM parameters.

Moreover, it was found in Ref.~\cite{Emmanuel-Costa:2016gdp} that the following pair of texture-zero mass matrices are also weak basis equivalent to an arbitrary $3 \times 3$ complex mass matrix:
\begin{align}
    M_u= \begin{pmatrix}
        0&\times&0\\
        \times&0&\times\\
        0&0&\times
    \end{pmatrix}, \qquad
   M_d= \begin{pmatrix}
        0&\times&0 \\
        \times&0&\times \\
        0&\times&\times
    \end{pmatrix}.
    \label{mat:costa}
\end{align}
Hence, (\ref{mat:d18}) and (\ref{mat:costa}) do not impose any constraints or relations on the quark masses or CKM parameters. Furthermore, replacing any of the texture-zeros with a free parameter will not spoil this freedom. Through appropriately replacing texture-zeros with free parameters and applying the permutations in (\ref{symm}) to (\ref{mat:d18}) and (\ref{mat:costa}), one can reproduce the texture-zero structures of all model classes in Table~\ref{tab:fulllist} except for $\mathbf{D_{1,8}}$, which was discussed above, and the already excluded $\mathbf{D_{3,3}}$. This proves that the remaining 17 equivalence classes of $\mN_{DW}=1$ models have sufficient freedom to reproduce the ten observables of the quark sector.

\section{Phenomenology of \texorpdfstring{$\mN_{DW}=1$}{NDW=1} variant models} 
\label{sec:pheno}

In this section we sketch some of the phenomenological processes which are common to many of the models in our classification. This section thus forms a point of departure for more in-depth studies of specific models in the future. 

\subsection{Flavour-changing Higgs processes}

In the SM, flavour-changing Higgs processes (FCHPs) arise at loop level and are suppressed. In 2HDM models, tree-level FCHPs arise when $\xi^q$ and $\eta^q$, as defined in \cref{eq:higgsbasisyukawa}, are not simultaneously diagonalizable. All model classes in Table~\ref{tab:fulllist} give rise to such processes.

These flavour-changing couplings induce tree-level contributions to meson mixing ($B_s$-$\bar{B}_s$, $B$-$\bar{B}$, $D$-$\bar{D}$ and $K$-$\bar{K}$), and rare meson decays such as $B_s\to \mu^+ \mu^-$. They also lead to flavour-violating Higgs decays such as $h\to bs$ or $h\to cu$. These cannot be effectively probed at the Large Hadron Collider; however, a future lepton collider such as FCC-ee may be able to improve on the mesonic constraints~\cite{Kamenik:2023ytu}.

Higgs flavour violation also leads to the top-quark decays $t \to ch$ and $t \to uh$ which are sensitive probes for new physics~\cite{ATLAS:2022gzn, Aguilar-Saavedra:2004mfd, Mandrik:2018yhe}. These have been studied in the context of the top-specific scenario~\cite{Chiang:2015cba,Chiang:2017fjr}, which belongs to our class $\mathbf{D_{1,1}}$.

Our models only include flavour violation in the quark sector; however, the more widely studied lepton-flavour violating Higgs decays such as $h\to \tau\mu$ could also be present depending on the details of the lepton sector. For an analysis of the model-independent constraints on flavour-violating couplings in the general type-III 2HDM, see Ref.~\cite{Crivellin:2013wna} and also~\cite{Arhrib:2015maa,Herrero-Garcia:2019mcy, Badziak:2021apn, DiLuzio:2023ndz}.

While the detailed phenomenology is model dependent, some general statements can be made about certain model classes. In model classes where the up or down sector has a block diagonal form (classes $\mathbf{{D_{1,5}}}$, $\mathbf{D_{1,7}}$, $\mathbf{D_{1,8}}$, $\mathbf{D_{3,1}}$, $\mathbf{D_{3,4}}$, $\mathbf{D_{3,5}}$), flavour mixing only occurs between two families of that sector at tree level. Hence, certain flavour changing processes are suppressed. Taking the representative model of class $\mathbf{D_{3,5}}$ in \cref{tab:fulllist} as an example, there are no $t \to u$ and $c \to u$ transitions at leading order. Note that, through the permutations (\ref{symm}), each of these classes contains a model in which any given pair of transitions in the up or down sector is suppressed.

An interesting feature of class $\mathbf{D_{1,8}}$ is that there are exactly ten parameters in the Yukawa matrices (in addition to $\tan\beta$). These are therefore fixed by the measured values of the quark masses and CKM parameters. The upshot is that all new physical processes which arise from the Yukawa sector of this model class are determined by known physical parameters. This class of models was studied in Ref.~\cite{Bjorkeroth:2018ipq}.

\subsection{Flavour-violating axion couplings}

The same flavour structure that leads to flavour-changing Higgs processes also generates flavour off-diagonal axion--quark couplings. This can be seen clearly in \eqref{yukinhiggs} after substituting in the parameterisation of the scalar fields in \eqref{eq:scalar-param}. It is customary to perform an axion-dependent rephasing of the fermion fields to the basis where the axion instead has (axial-)vector couplings,
\begin{equation}
    \mathcal{L}_a\supset-\frac{\partial_{\mu}a}{2f_a}\left[\bar{f}_i\gamma^{\mu}\left(C^V_{f_if_j}-C^A_{f_if_j}\gamma_5 \right)f_j  \right] \,,
    \label{ainteract}
\end{equation}
with
\begin{equation}
    C_{f_if_j}^{V,A}=\frac{1}{2N}\left(U_L^{f\dagger}\boldsymbol{X}_{f_L}U_L^f \pm V_R^{f\dagger}(-\boldsymbol{X}_{f_R})V_R^f \right)_{ij}. \label{axioncoupling} 
\end{equation}
Here, $2N=\pm 1$, $U^f_L$ and $V^f_R$ are the unitary matrices that diagonalise the quark mass matrices, and in the second line $+$ ($-$) corresponds to the vector (axial-vector) coupling. The $\boldsymbol{X}_{f_{L,R}}$ are the diagonal, model-dependent matrices of PQ charges (see Table \ref{PQcharges}).\footnote{The sign of $\boldsymbol{X}_{f_R}$ in \eqref{axioncoupling} arises from the action of the PQ symmetry on the right-handed fields (\ref{pqsymm}).} Flavour-violating axion couplings then arise due to the flavour non-universality of the PQ charges. An explicit relation between the couplings $C_{f_if_j}^{V,A}$ and the flavour-violating couplings in the Higgs sector, $\xi^q$, is given in Ref.~\cite{DiLuzio:2023ndz}.

Meson decays provide a powerful probe of these flavour-violating axion interactions. The strongest experimental constraint on the $C_{f_if_j}^{V,A}$ parameters comes from the decay $K^+ \to \pi^+ a$, which gives the bound $f_a(\textrm{GeV})>8.3 \times 10^{11}|C_{sd}^V|$~\cite{Bauer:2021mvw,NA62:2021zjw}. Detailed analyses of the bounds on flavour-violating axion couplings can be found in Refs.~\cite{MartinCamalich:2020dfe,Bauer:2021mvw}. 

The structure of the couplings $C^{V,A}_{f_if_j}$ is model-dependent. However, since the source of axion flavour-violation is the same as the source of Higgs flavour-violation, the comments made in the previous section regarding certain model classes also apply to the axion-quark interactions. Namely, model classes which exhibit a block diagonal form for either the up or down sector Yukawas ($\mathbf{{D_{1,5}}}$, $\mathbf{D_{1,7}}$, $\mathbf{D_{1,8}}$, $\mathbf{D_{3,1}}$, $\mathbf{D_{3,4}}$, $\mathbf{D_{3,5}}$), will induce flavour-violation solely between two families in the given sector. In other words, $C^{V,A}_{f_if_j}$ is block diagonal for that sector. Finally, in class $\mathbf{D_{1,8}}$ models, the axion--quark couplings are fixed in terms of measured observables (up to factors of $\tan\beta$), in analogy with the flavour-changing Higgs couplings of this class.

\subsection{Nucleophobic axions}

Flavour-variant DFSZ models have the potential to suppress the axion-nucleon coupling, making the axion ``nucleophobic''. This concept was first considered in Ref.~\cite{Hindmarsh:1997ac} by coupling the axion only to the up quark. Recently, there has been a resurgence of interest in nucleophobic axions, motivated by relaxing the upper bound on the axion mass $m_a\lesssim~20\textrm{ meV}$ from the neutrino burst of SN1987A~\cite{Carenza:2019pxu}. 

For DFSZ-type models, such a suppression can only be arranged for $\mN_{DW}=1$ flavour-variants~\cite{DiLuzio:2017ogq}. This requires the colour anomaly to be completely determined by the first generation of quarks, in addition to a fine tuning of model parameters.\footnote{The inclusion of three Higgs doublets may allow nucleophobia without fine tuning or $\mN_{DW}=1$~\cite{Badziak:2023fsc}.} Each equivalence class in Table~\ref{tab:fulllist} contains a model that can be engineered to exhibit nucleophobia.

\section{Conclusion} 
\label{sec:conclusion}

The DFSZ model is a well-motivated scenario that provides a solution to the strong CP problem, and a viable dark matter candidate in the axion. Nevertheless, it suffers from a well-known cosmological domain wall problem. Flavoured DFSZ models with unit domain wall number provide a robust and elegant solution to this problem, while simultaneously giving rise to an intriguing connection between the solution to the strong CP problem and the broader flavour puzzle of the Standard Model.

In this work, we have systematically identified all possible three-family, flavoured DFSZ models free of a domain wall problem, thus uncovering a catalogue of variant DFSZ models which have the prospect of consistent and complete cosmological histories, as per Ref.~\cite{Sopov:2022bog} for the top-specific case.

Many of these models have not previously been identified in the literature. They exhibit a rich flavour structure that interconnects axion and flavour physics. They generically feature texture-zeros in the Yukawa sector that have an underlying symmetry justification due to the PQ symmetry.

We have classified the viable models into 17 equivalence classes under permutations of the fields \eqref{symm}. They are summarised in Table~\ref{tab:fulllist}, with their corresponding PQ charge structures in Table~\ref{PQcharges}. Each equivalence class contains a large number of phenomenologically distinct models, with different flavour-violating axion and Higgs interactions.

These models produce a diverse range of phenomenological signatures, with effects in flavour-violating Higgs decays, meson mixing, and rare meson decays. The flavour-violating axion interactions provide a complementary probe of axion interactions beyond the standard axion processes, and could serve to distinguish these models from a standard DFSZ or KSVZ axion. 

It would be interesting to explore these models in further detail, including extensions that incorporate the lepton sector. Since leptonic PQ charges do not contribute to the colour anomaly, there is significant freedom in the flavour structures that could be considered. This would further enrich the phenomenology of these models.

\acknowledgments
This work was supported in part by the Australian Research Council through the ARC Centre of Excellence for Dark Matter Particle Physics, CE200100008. RRV was also partially supported by the Australian Research Council Discovery Project DP200101470. AT was partially supported by the Australian Research Council Discovery Project DP210101900. PC is supported by the Australian Research Council Discovery Early Career Researcher Award DE210100446.

\appendix


\bibliographystyle{JHEP}
\bibliography{biblio.bib}

\end{document}